\documentclass[apjl]{emulateapj}
\usepackage{natbib}
\usepackage{graphicx}
\usepackage[usenames]{color}
\usepackage{wasysym}

\newcommand{\Deg}{\ensuremath{^\circ}}

\shorttitle{Central rotations of Milky Way Globular Clusters}
\shortauthors{Fabricius et al.}

\begin{document}
\title{Central rotations of Milky Way Globular Clusters$^{\dagger}$}

\author{
Maximilian H. Fabricius,\altaffilmark{1,2}
Eva Noyola,\altaffilmark{3}
Surangkhana Rukdee,\altaffilmark{1}
Roberto P. Saglia,\altaffilmark{1}
Ralf~Bender,\altaffilmark{1,2}
Ulrich Hopp,\altaffilmark{1,2}
Jens~Thomas,\altaffilmark{1}
Michael Opitsch,\altaffilmark{2,4}
Michael J. Williams\altaffilmark{1,5}
}

\altaffiltext{1}{Max Planck Institute for Extraterrestrial Physics,
Giessenbachstrasse, 85748 Garching, Germany}
\altaffiltext{2}{University Observatory Munich, Scheinerstra\ss e 1, 81679
Munich, Germany}
\altaffiltext{3}{McDonald Observatory, The University of Texas at Austin, 
2515 Speedway, Stop C1402, Austin, Texas 78712-1206, USA}
\altaffiltext{4}{Excellence Cluster Universe, Boltzmannstr. 2, D-85748 Garching, Germany}
\altaffiltext{5}{Department of Astronomy, Columbia University, New York 10027, USA}

\begin{abstract} 
Most Milky Way globular clusters (GCs) exhibit measurable flattening, even if
on a very low level. Both cluster rotation and tidal fields are thought to
cause this flattening. Nevertheless, rotation has only been confirmed in a
handful of GCs, based mostly on individual radial velocities at large radii.
We are conducting a survey of the central kinematics of Galactic GCs using the
new Integral Field Unit instrument VIRUS-W. We detect rotation in all 11 GCs
that we have observed so far, rendering it likely that a large majority of the
Milky Way GCs rotate. We use published catalogs of the ACS survey of GCs to
derive central ellipticities and position angles. We show that in all cases
where the central ellipticity permits an accurate measurement of the position
angle, those angles are in excellent agreement with the kinematic position
angles that we derive from the VIRUS-W velocity fields. We find an unexpected
tight correlation between central rotation and outer ellipticity, indicating
that rotation drives flattening for the objects in our sample. We also find a
tight correlation between central rotation and published values for the central
velocity dispersion, most likely due to rotation impacting the old dispersion
measurements.

\vspace{.25 cm} 
\noindent {}$^{\dagger}${This paper includes data taken at The McDonald
Observatory of The University of Texas at Austin.} \end{abstract}


\section{Introduction}

Globular Clusters (GCs) have historically been viewed as simple systems whose
evolutionary history is well understood, but new observations keep revealing
surprising results. GCs have short central relaxation times compared to their
ages. Anisotropies and rotation are therefore likely to be very small in the
central regions, while a moderate amount is expected in the outskirts
\citep{Meylan1997}.

Flattening (with a median axial ratio $\approx$ 0.9) is found in the outer
regions of Galactic GCs \citep{White1987,Chen2010}. Clusters closer to the
bulge tend to be more flattened than those in the halo. Given this observed
flattening, rotation is expected in the outer parts of GCs, and indeed it has
been measured for a number of them. Table\,7.2 in \citet{Meylan1997} includes
all the relevant references before 1997. More recent work by \citet{Lane2011}
and \citet{Bellazzini2012} and references in Tables 1 \& 2 of \citet{Zocchi2012} use
individual radial velocity measurements to show that anisotropy increases
towards the outskirts of clusters. It is still under debate how much of the
observed flattening seen for the outer parts of GCs is due to galactic tidal
effects, disk cross shocking or to rotation.

On the theory side, analytical \citep{Longaretti1997}, Fokker-Planck
\citep{Einsel1999}, and $N$-body models \citep{Ernst2007} indicate that the
presence of rotation affects the dynamical evolution of single mass star
clusters, by accelerating core-collapse time scales. The caveat is that the
effect seems to vanish for isolated two-mass $N$-body models \citep{Ernst2007}.
Recent models indicate that rotation could be a key ingredient in the formation
of multiple generations of stars in GCs
\citep{Bekki2010,Mastrobuono-Battisti2013}. The time evolution of $N$-body
models seems to indicate that the rotation signature for the central region
gets erased after a few relaxation times (Fig.\,14 of \citealp{Ernst2007}),
which implies that rotation is not expected around the core radius of relaxed
GCs.

Detailed modelling including central kinematics has been performed for a few
clusters. In particular, $\omega$\ Cen \citep{van-Leeuwen2002}, 47\ Tuc
\citep{Anderson2003} and M15 \citep{van-den-Bosch2006} have been observed to
rotate using various datasets and analysis techniques. \citet{Bianchini2013}
undertakes the most thorough dynamical modelling for these three clusters to
date. They find that for all three clusters, which have very different
dynamical states, the ellipticity and anisotropy decrease towards the center,
to the point of being practically isotropic inside the core radius. Besides
these three cases, kinematic observations covering the regions around one or
two core radii are scarce.

We are collecting optical, high resolution Integral Field Unit (IFU)
spectroscopy of Milky Way GCs in an ongoing survey. In this letter we present
data for the first 11 clusters that we have observed.

\begin{figure*}
\begin{center}
\includegraphics[width=0.3\textwidth]{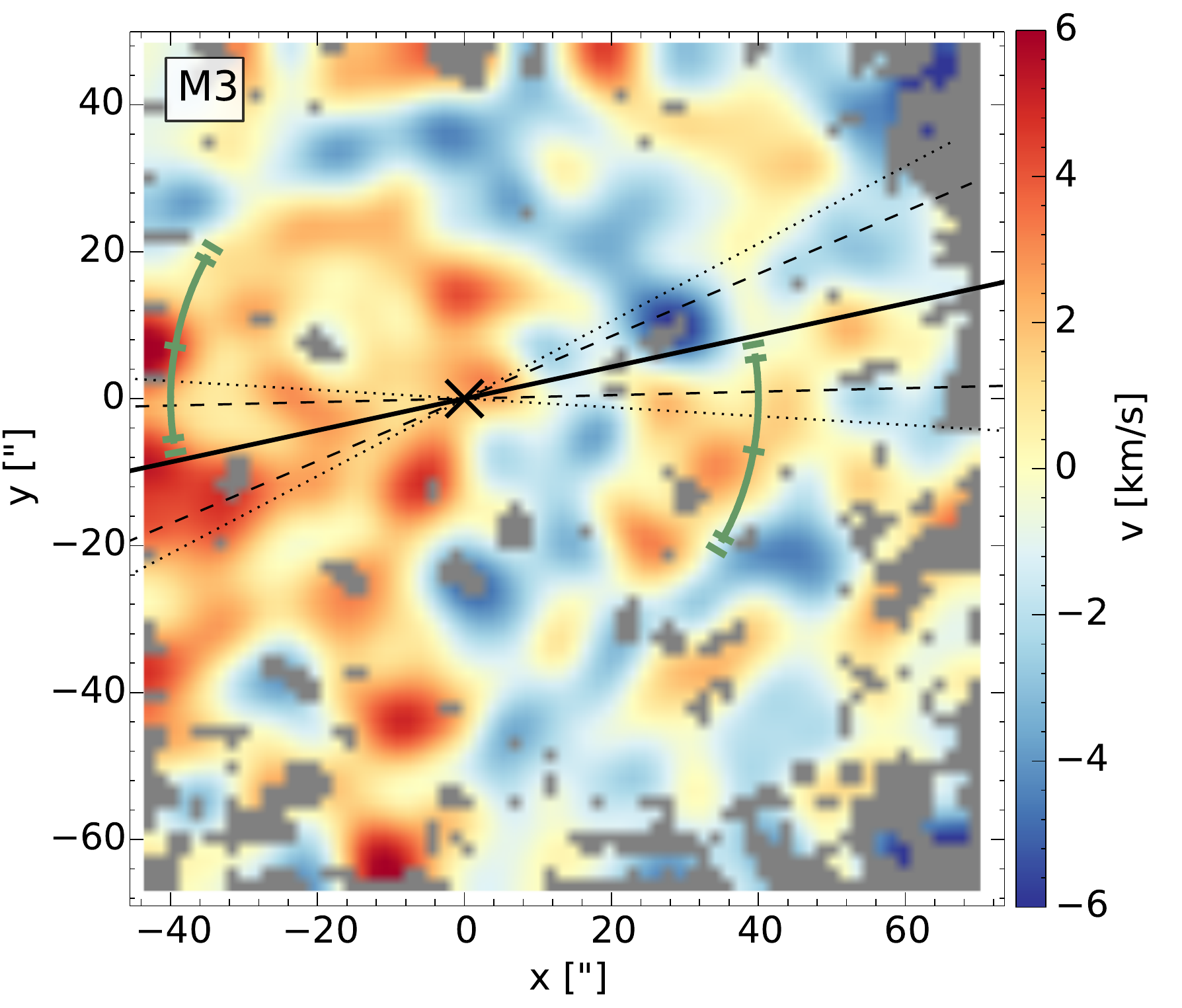}
\includegraphics[width=0.3\textwidth]{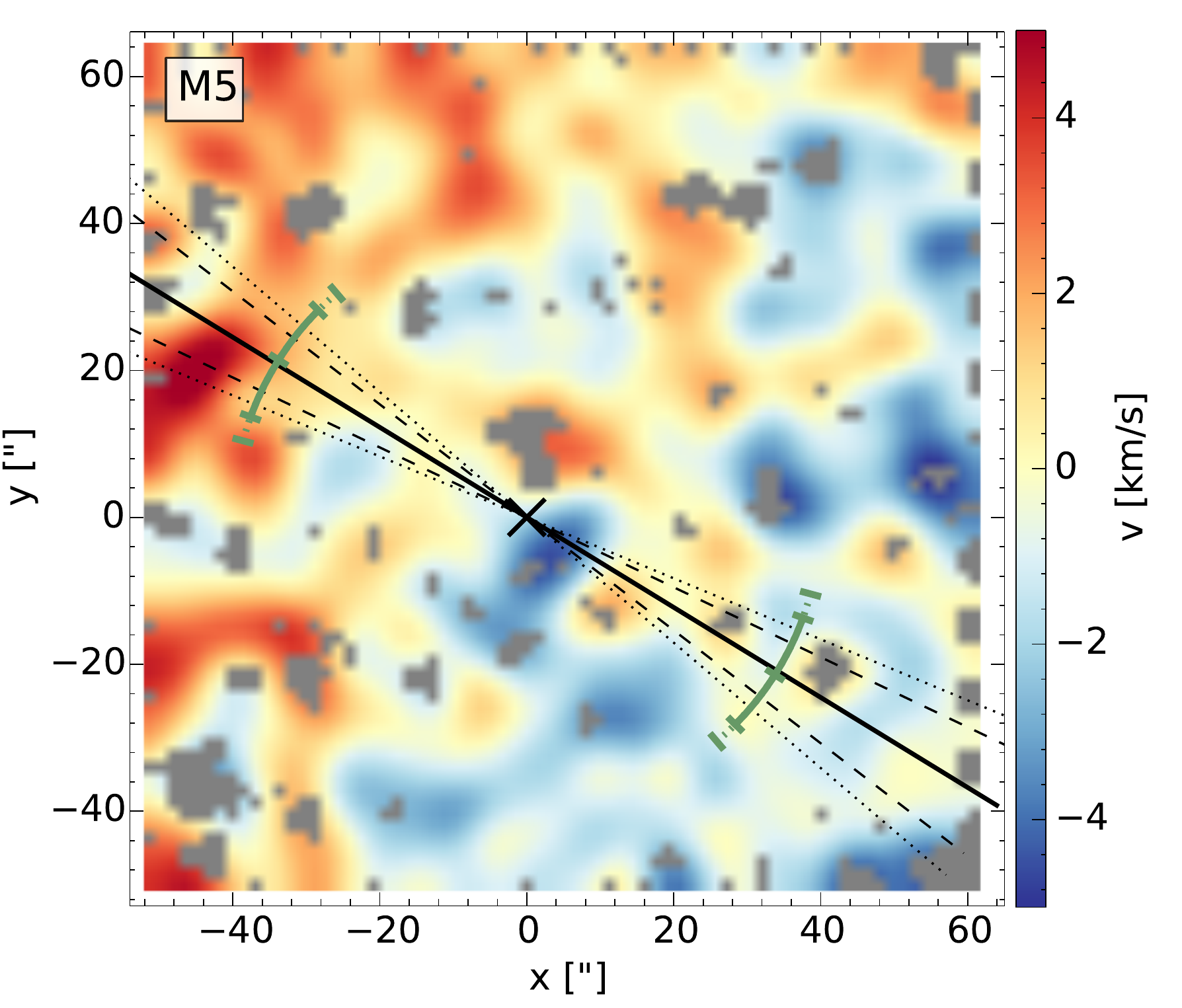}
\includegraphics[width=0.3\textwidth]{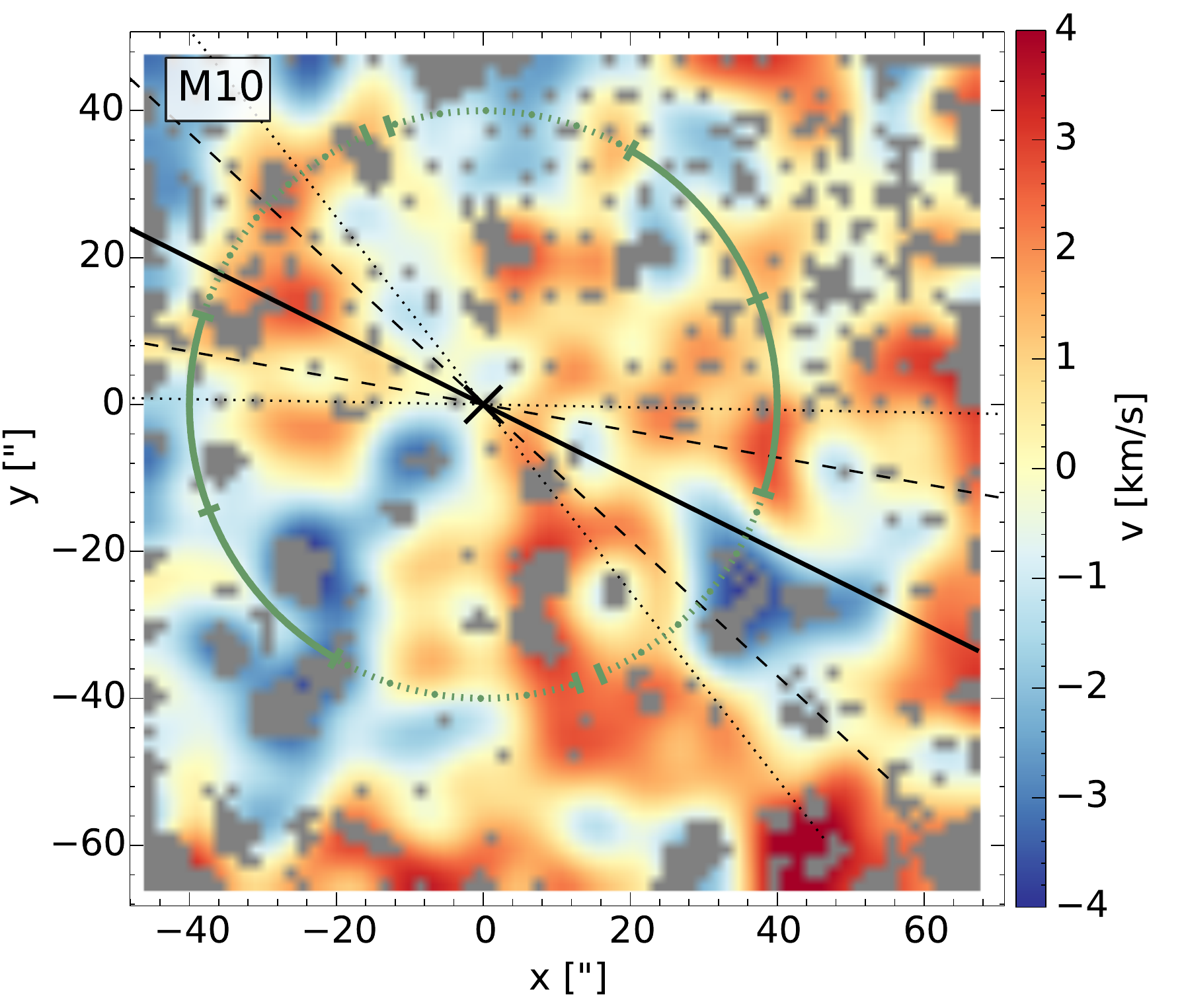}\\
\includegraphics[width=0.3\textwidth]{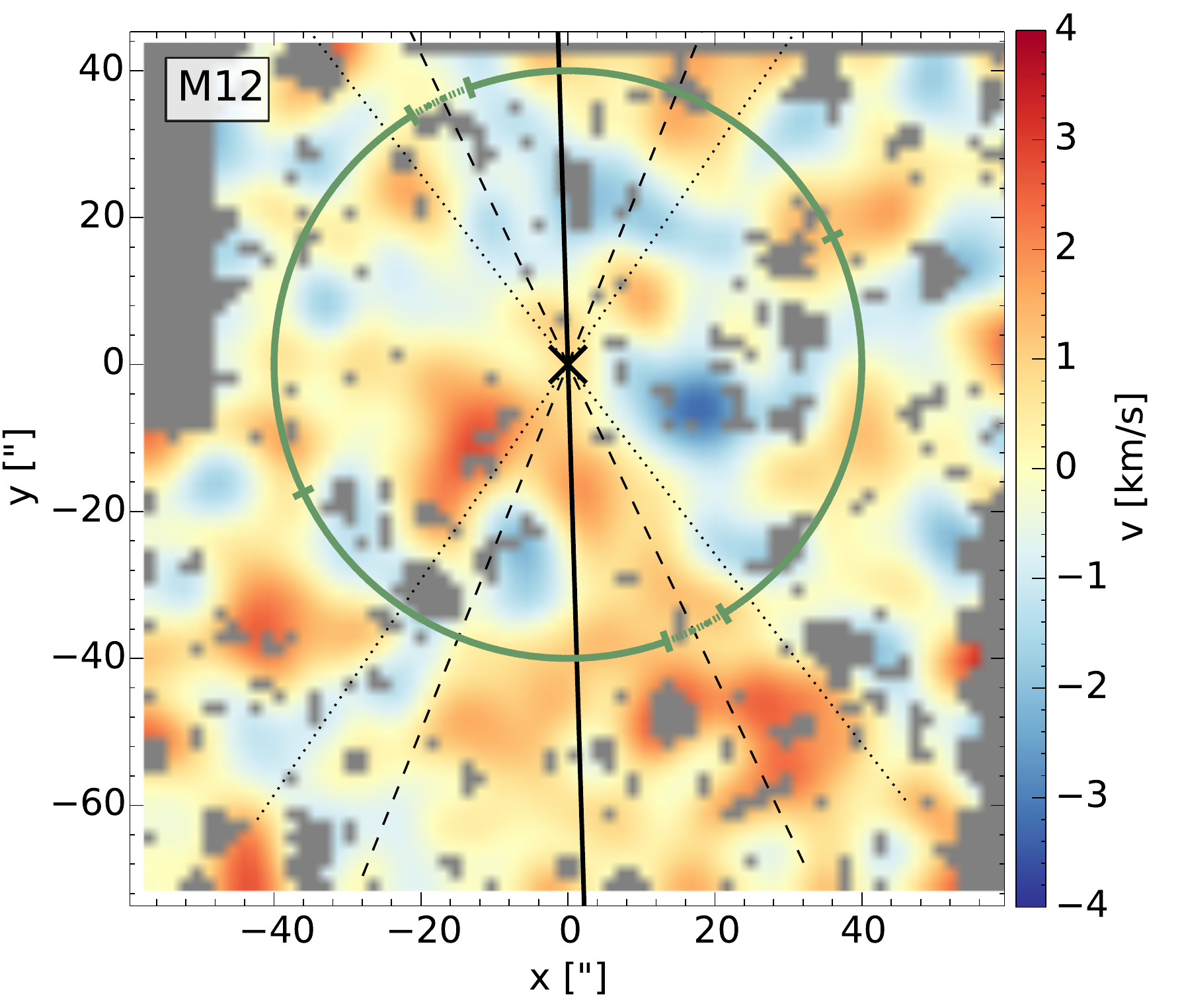}
\includegraphics[width=0.3\textwidth]{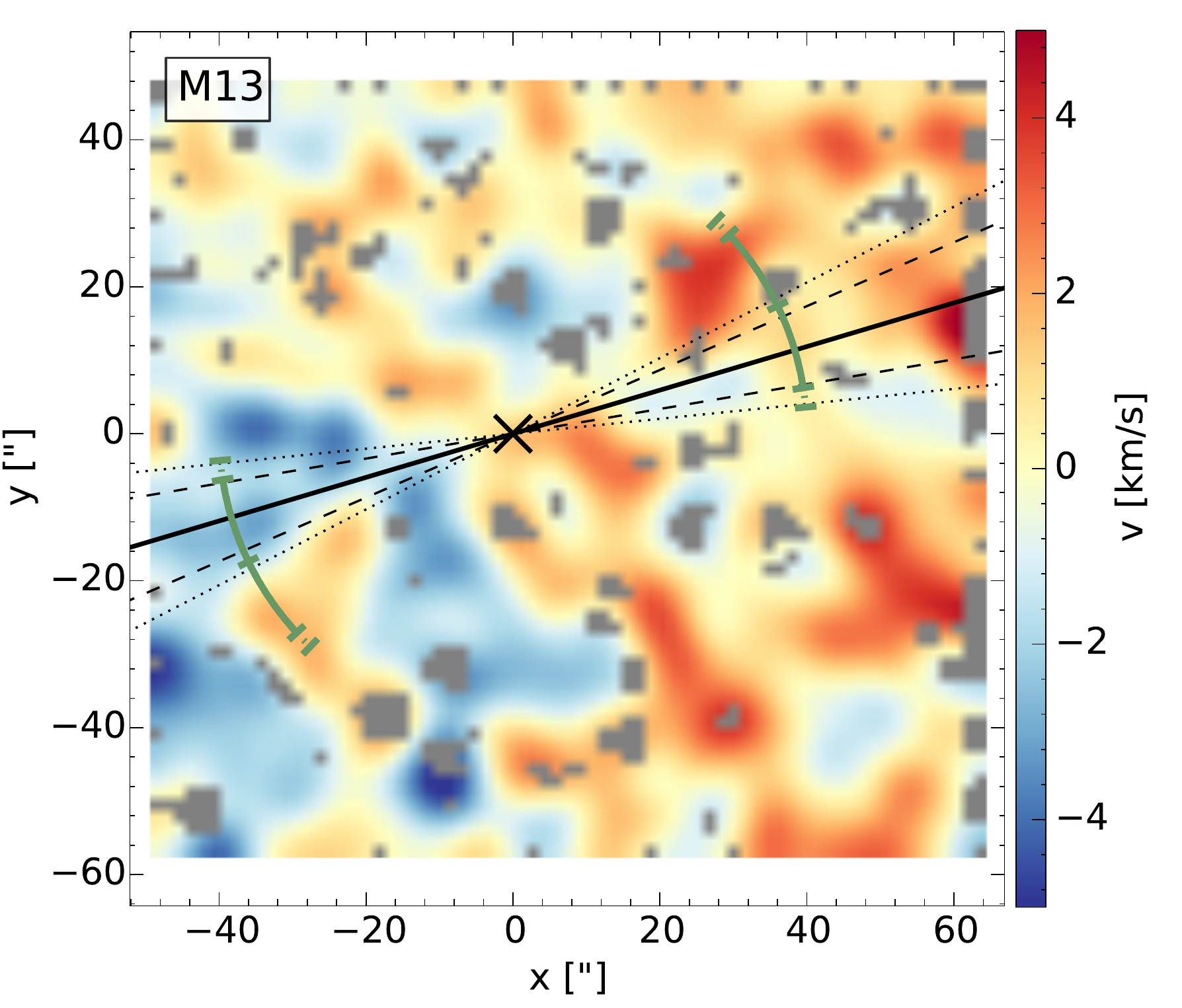}
\includegraphics[width=0.3\textwidth]{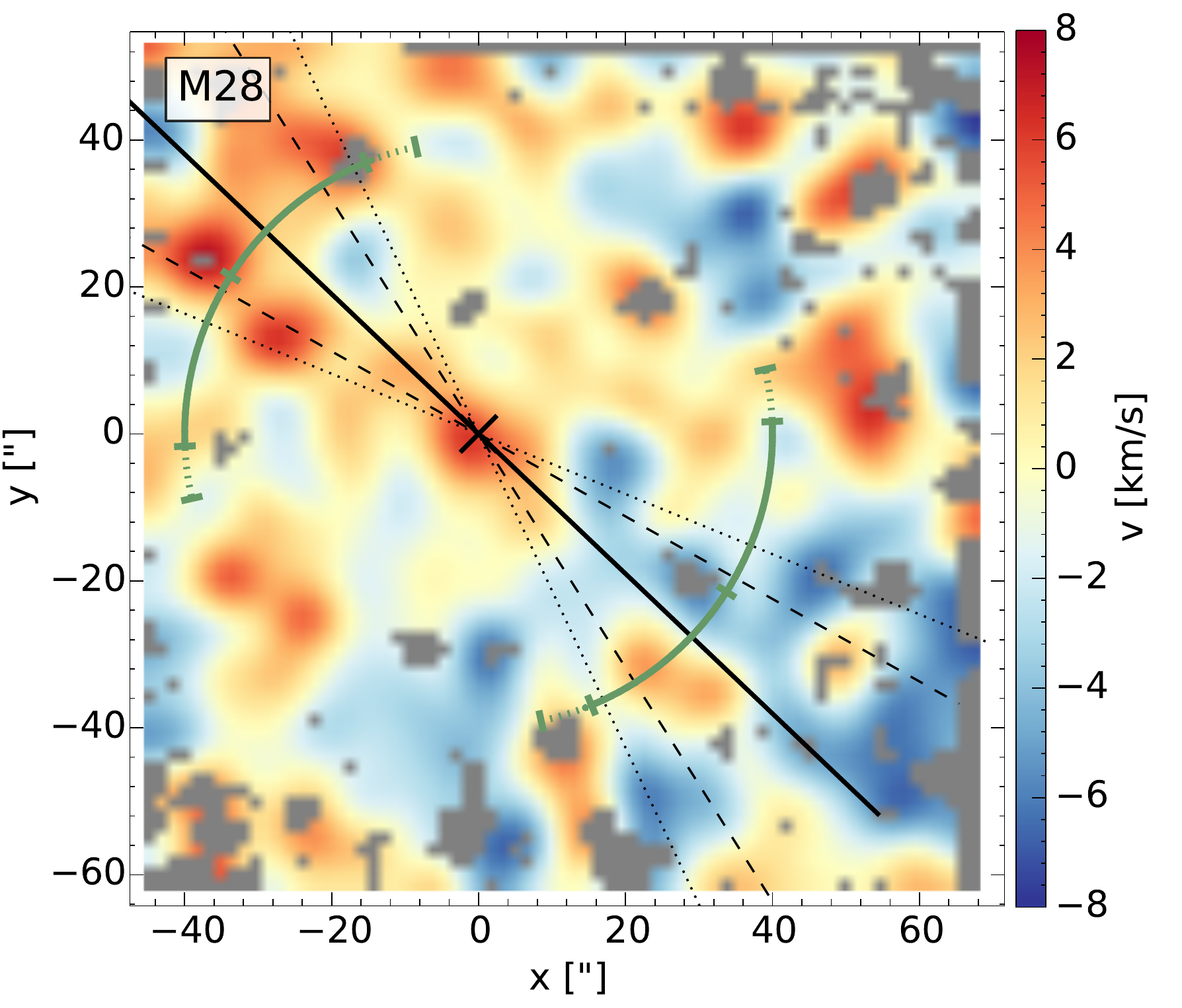}\\
\includegraphics[width=0.3\textwidth]{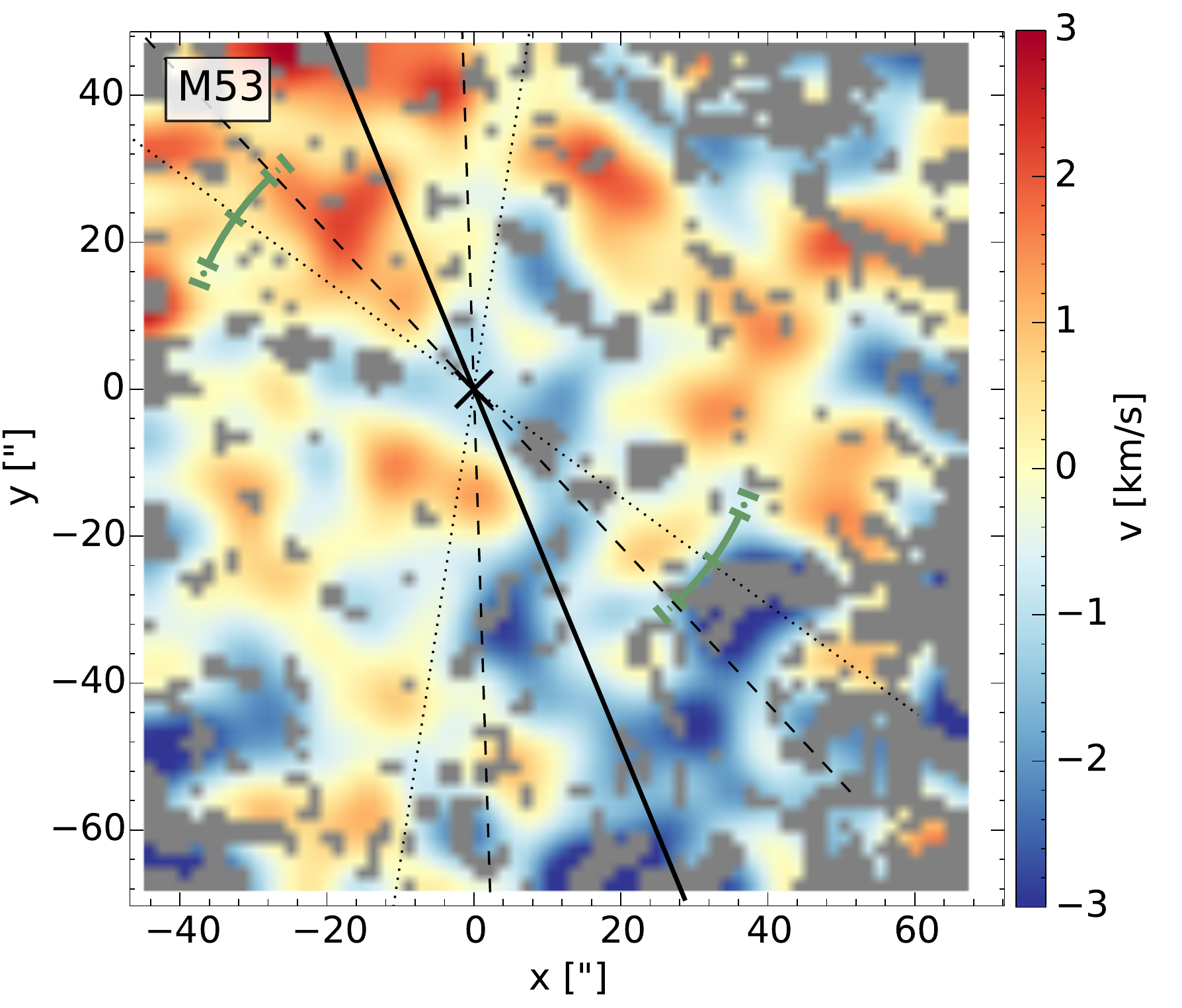}
\includegraphics[width=0.3\textwidth]{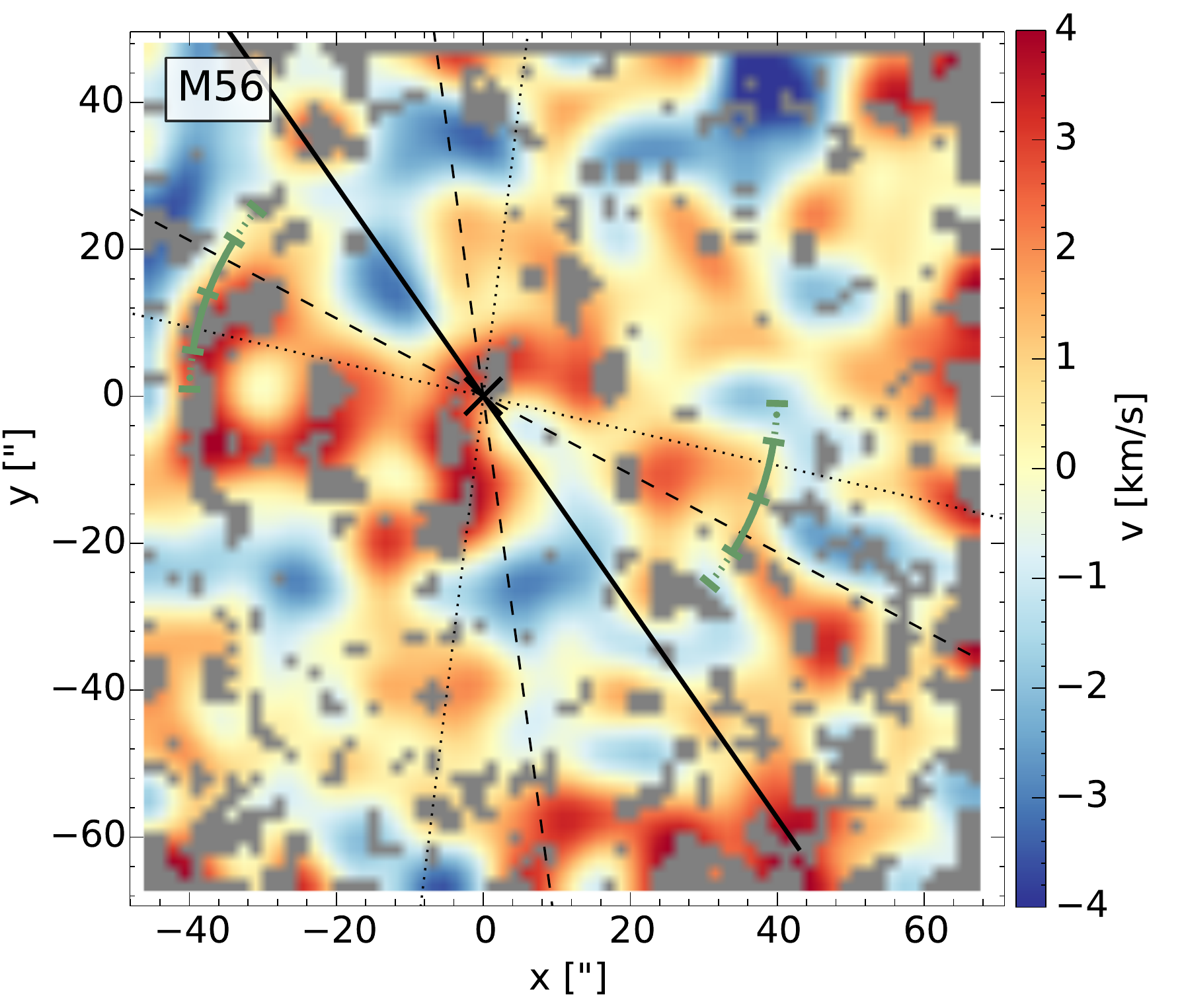}
\includegraphics[width=0.3\textwidth]{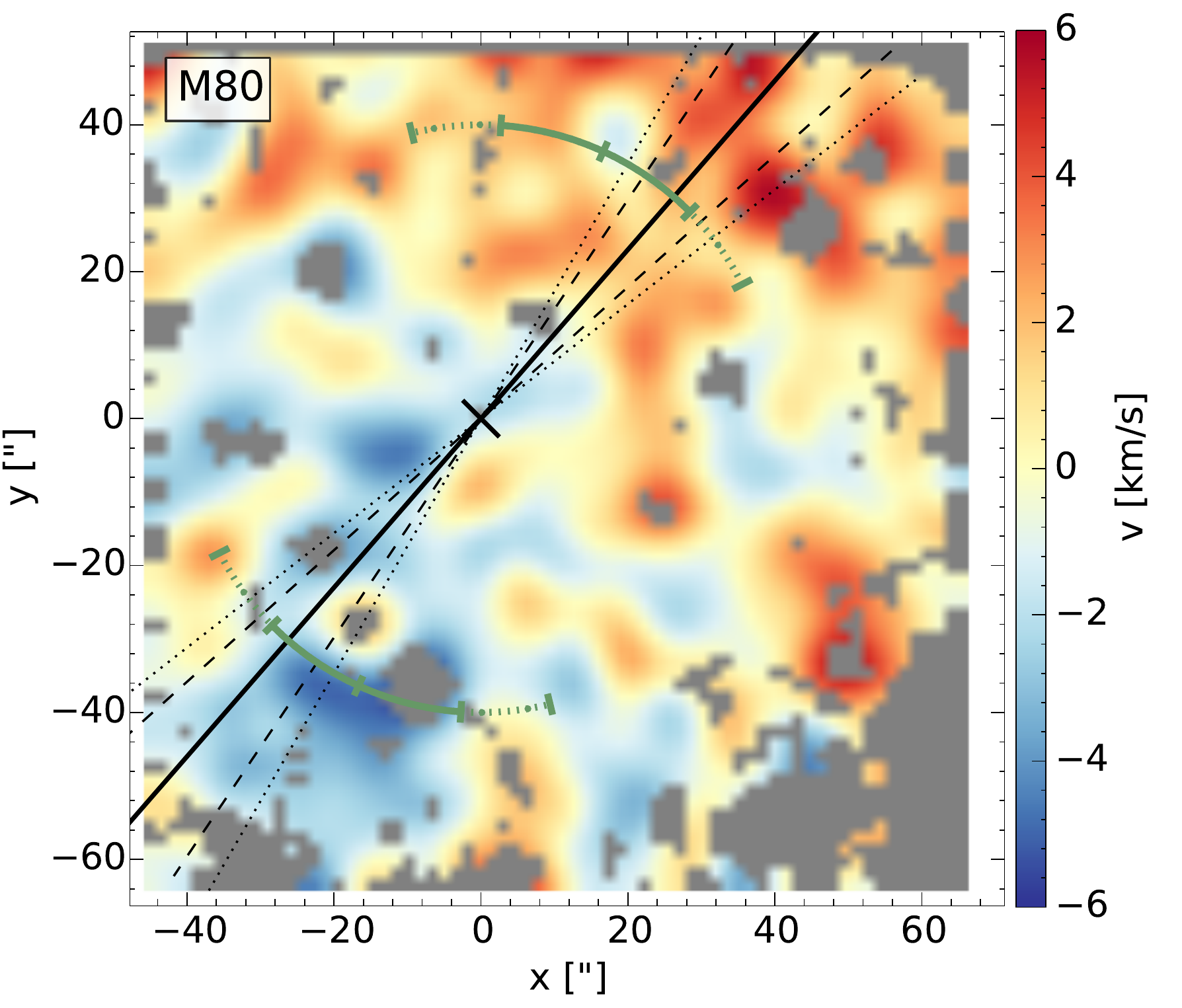}\\
\includegraphics[width=0.3\textwidth]{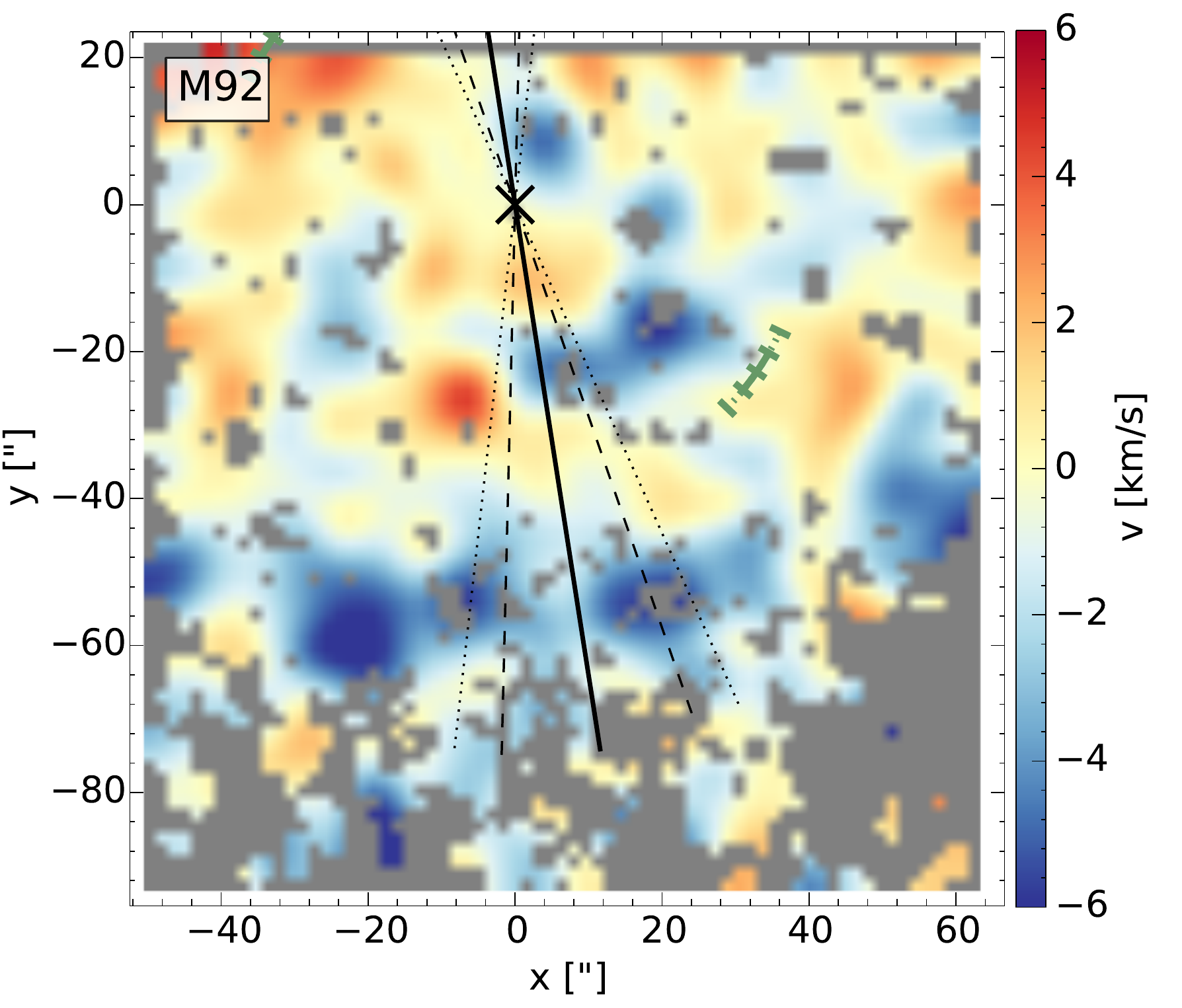}
\includegraphics[width=0.3\textwidth]{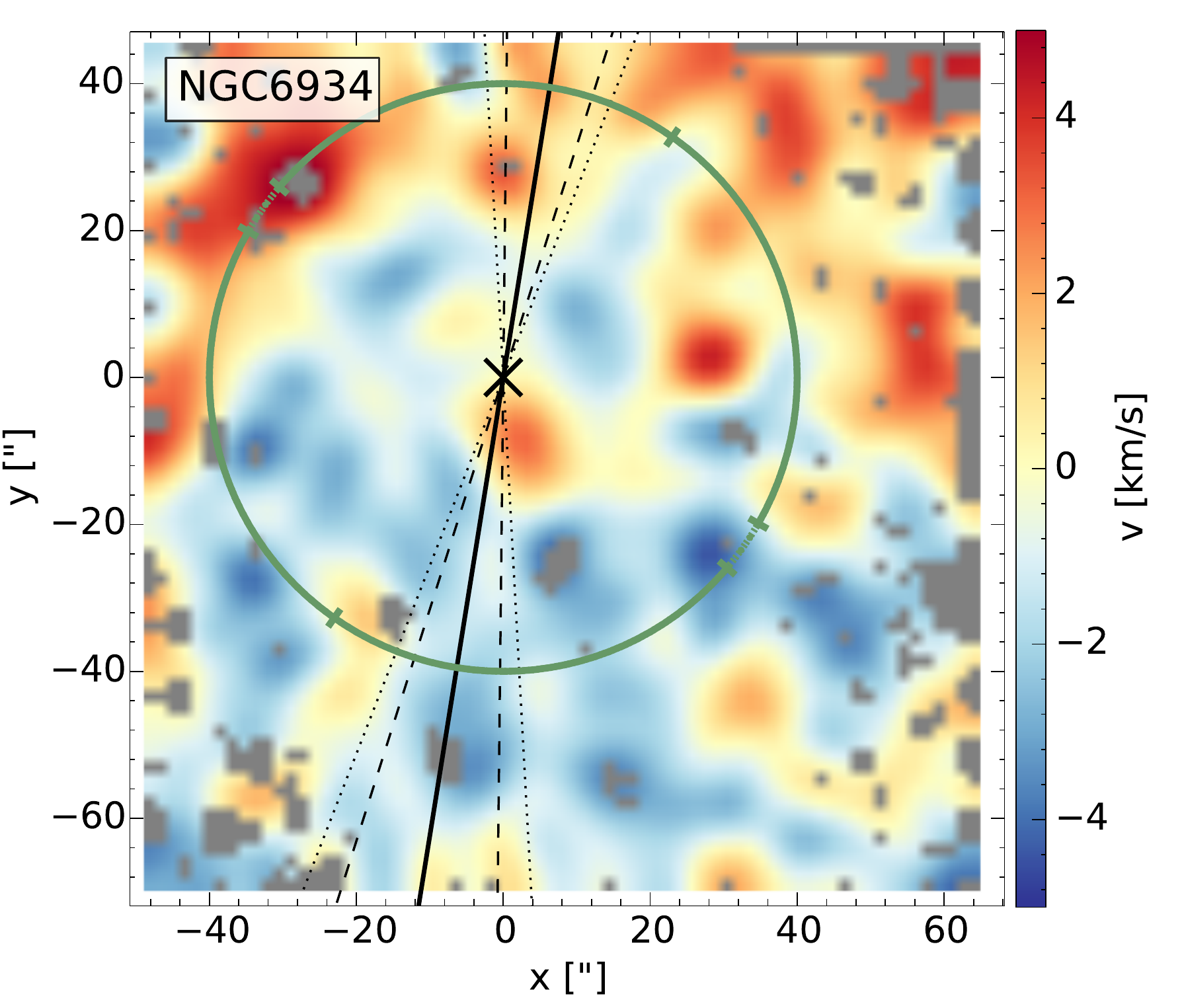}
\end{center}
\caption{
Velocity fields for the 11 GCs in our sample. The plots show the velocities
that we derive on a per-pixel basis. The maps are smoothed with a Gaussian
kernel with a width of 2 pixels to highlight the rotation signature.  Grey
areas indicate spectral pixels that were rejected from the kinematic analysis
based either on our signal-to-noise cut of 5 or the kappa-sigma clipping
process. The cross indicates the cluster center, while the straight solid line
shows the kinematic position angle. The dashed line shows the systematic
uncertainty and the dotted line shows the systematic plus the statistical
uncertainty. The green arcs indicate the eigenvalue derived photometric
position angle. The solid part of the arc represents the systematic uncertainty
and the dotted part the systematic plus the statistical uncertainty. In all
plots the $x$ axis is aligned with the east-west direction and positive $x$
values point west while $y$ increases towards the north. }
\label{fig:vmaps_ellmaps}
\end{figure*}

\section{Sample \& Observations}
\label{sec:observations}

Our sample was constructed based only on feasibility: we first selected all
objects that are observable from the McDonald Observatory (at airmass $<$ 2)
from the catalog of Milky Way GCs by \citet{Harris1996} (hereafter H96) and for
which surface brightness estimates \citep{Trager1995, Noyola2006} led to
reasonable exposure times ($< 6$\,h). This gave a sample of 27 clusters, 11 of
which are presented here.

The observations were carried out  during August 2012, and April and August
2013 using the fiber-based IFU Spectrograph VIRUS-W \citep{Fabricius2012b} at
the 2.7\,m Harlan J.\ Smith Telescope of the McDonald Observatory in Texas. We
used the higher resolution mode of the instrument with $R \sim 9000$ and a
wavelength coverage of 4855\,\AA -- 5475\,\AA. The IFU has a field of view of
105\arcsec$~\times$~55\arcsec\ with the long edge always aligned with the
east-west axis. The fibers are 3.2\arcsec\ in diameter on sky and are arranged
in a dense-pack configuration with a fill-factor of 1/3 such that three
dithered observations fill in the gaps between fibers.

For each cluster we observed two offset but slightly overlapping pointings such
that the combined field of view amounts to about
105\arcsec$~\times$~105\arcsec. Depending on surface brightness, we took 600\,s
to 2400\,s exposures in each dither position (see Table\,\ref{tab:results}),
split in half for cosmic ray rejection. We took 600s empty sky exposures
between each dither position and recorded bias frames and Hg and Ne arc lamp
exposures for calibration every night.

\section{Data reduction and kinematic extraction}
\label{sec:data_reduction}

Our data reduction uses the {\tt fitstools} package by \citet{Gossl2002} and
the {\tt Cure} pipeline developed by our group for HETDEX \citep{Hill2004}.
The procedure follows standard prescriptions for the generation of master
biases, flats and arc lamp frames.
 
After the spectral extraction, we average the two cosmic ray split exposures and
the bracketing sky exposures, rejecting spurious events. We scale the sky spectra by
exposure time and subtract them from the science exposures. We combine the
individual fiber spectra of the two dithered pointings into one datacube by
imposing a pixel grid with an edge length of 1.6\arcsec. A detailed description
of the reduction is given in \citet{Fabricius2014}.

Given the 3.2\arcsec\ fibers, we typically integrate over the light of several
stars per fiber. A simple cross correlation method may therefore yield
uncertain velocities. We use a newly implemented version of the Maximum
Penalized Likelihood Method by \citet{Gebhardt2000b} (see
\citealp{Fabricius2014}) to first recover non-parametric line of sight velocity
distributions (LOSVD). We then fit Gaussian models to each LOSVD while only
taking velocity channels that are separated by no more than 40\,km\,s$^{-1}$
from the systemic velocity into account. We extract the ten brightest spectra
of each datacube to use as templates, since they are most likely dominated by
the light of individual bright stars. We pick the spectrum that delivers the
lowest RMS of the residuals in the non-parametric fit as final template. We
employ no spatial binning but reject spectra with a mean signal-to-noise below
5. Typical errors for the recovered mean velocities are 1.5\,km\,s$^{-1}$.
 
The derivation of velocities for each pixel produces a velocity field to which
we fit a plane parametrized by
\begin{eqnarray}
 v(x,y) = a x + b y + v_{\mathrm{sys}} 
\end{eqnarray}
where the slopes $a$ and $b$ and the systemic velocity $v_{\mathrm{sys}}$ are free
parameters.  The fit is carried out using a standard minimum least square
fitting routine (MINPACK {\tt lmdif}; \citealp{More1980}). From this, we obtain
a kinematic position angle $PA_{\mathrm{kin}} = \mathrm{arctan}(b/a)$ and the absolute value of
the central velocity gradient $\|\nabla v\| = \sqrt{a^2 + b^2}$. During the
fit, we reject outlier pixels (in velocity) through $\kappa-\sigma$ clipping with
$\kappa = 2.5$. The mean RMS of the residuals of the plane fit is
3.3\,km\,s$^{-1}$ over all GCs. We estimate the statistical uncertainty through
bootstrapping: we draw one hundred subsamples with replacement and perform the
plane fit for each of the subsamples. Individual pixels in the velocity fields
are correlated over scales of up to five pixels because of fiber size and
seeing. The bootstrapping method therefore does not reflect the systematic
uncertainty caused by a few high-velocity stars. To assess this effect, we
generate another 100 datasets where we mask a total of 11 randomly placed
7\,px~$\times$~7\,px, subregions, covering about 10\,\% of all pixels. We
repeat the plane fit to all these 100 datasets and report the total spread of
position angles and gradients as systematic uncertainty.

\section{Photometric position angles and central ellipticities}

In order to test if the observed rotation is reflected in the morphology, we
derive ellipticities and position angles using catalogs from the HST ACS survey
of GCs \citep{Sarajedini2007} that are available for 10 out of our 11 clusters.
We obtain the ellipticities and position angles from an eigenvector analysis of
the spatial star distribution. For this we compute the $2 \times 2$ covariance
matrix of the stellar positions as
\begin{eqnarray}
	V_{x,y} = \mathrm{cov}(x,y) = \langle (x - \mu_x)(y - \mu_y) \rangle
\end{eqnarray}
where $x$ and $y$ are the catalog coordinates and $\mu_x$
and $\mu_y$ are the adopted center coordinates of the cluster. 
The eigenvectors $v_1, v_2$ and the eigenvalues $\lambda_1,
\lambda_2$ then yield the ellipticity and the position angle through
$\varepsilon = 1 - \sqrt{\lambda_2/\lambda_1}$ and $PA =
\mathrm{arctan}(v_{1,y}/v_{1,x})$. We reject stars outside a radius of
100\arcsec\, which reaches the edge of the field of view of ACS-WFC. The
catalogs exclude non cluster-members.

We compute the center positions iteratively by shifting the $r = 100$\arcsec\
aperture until the first moments in both spatial directions become zero with
respect to the new center position. We compare these centers with the ones
derived by \citet{Noyola2006} for the common objects. The differences are under
2\arcsec\ on average. We tested that the aperture truncation does not affect our
reported ellipticities. Using radial bin counts, we have further tested that
the derived ellipticities and position angles are not caused by artefacts in
the catalogs such as lower star counts around bright stars. 

In general, our central ellipticities are small compared to published values at
larger radii by up to a factor of ten (compared for instance with
\citealp{Geyer1983,White1987,Chen2010}). Also, we find no obvious correlation
between our results and published ellipticities. This however is easily
explained by the fact that our data probe vastly different radial scales.
 
\begin{figure*}
\begin{center}
\includegraphics[width=0.4\textwidth]{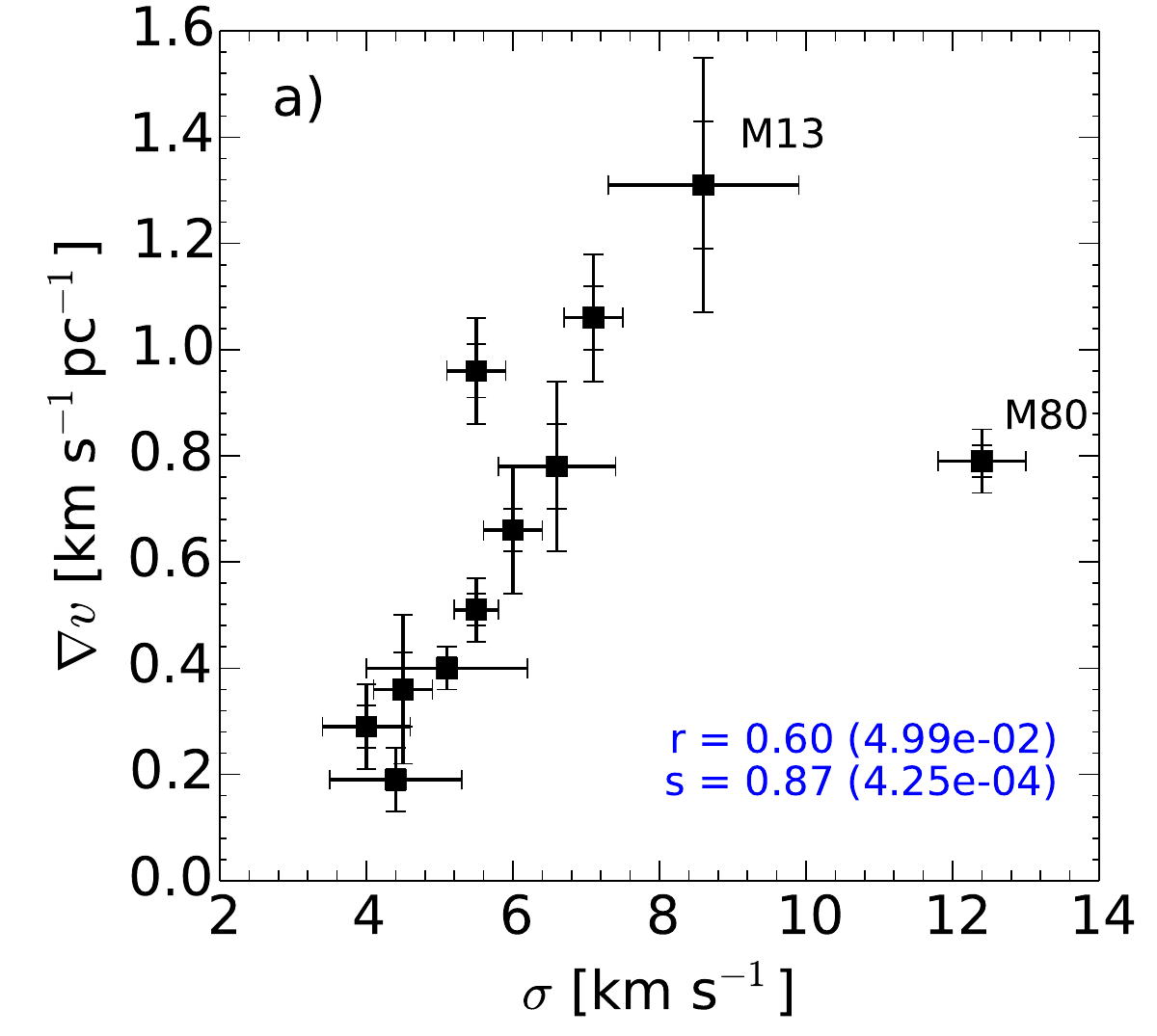}
\includegraphics[width=0.4\textwidth]{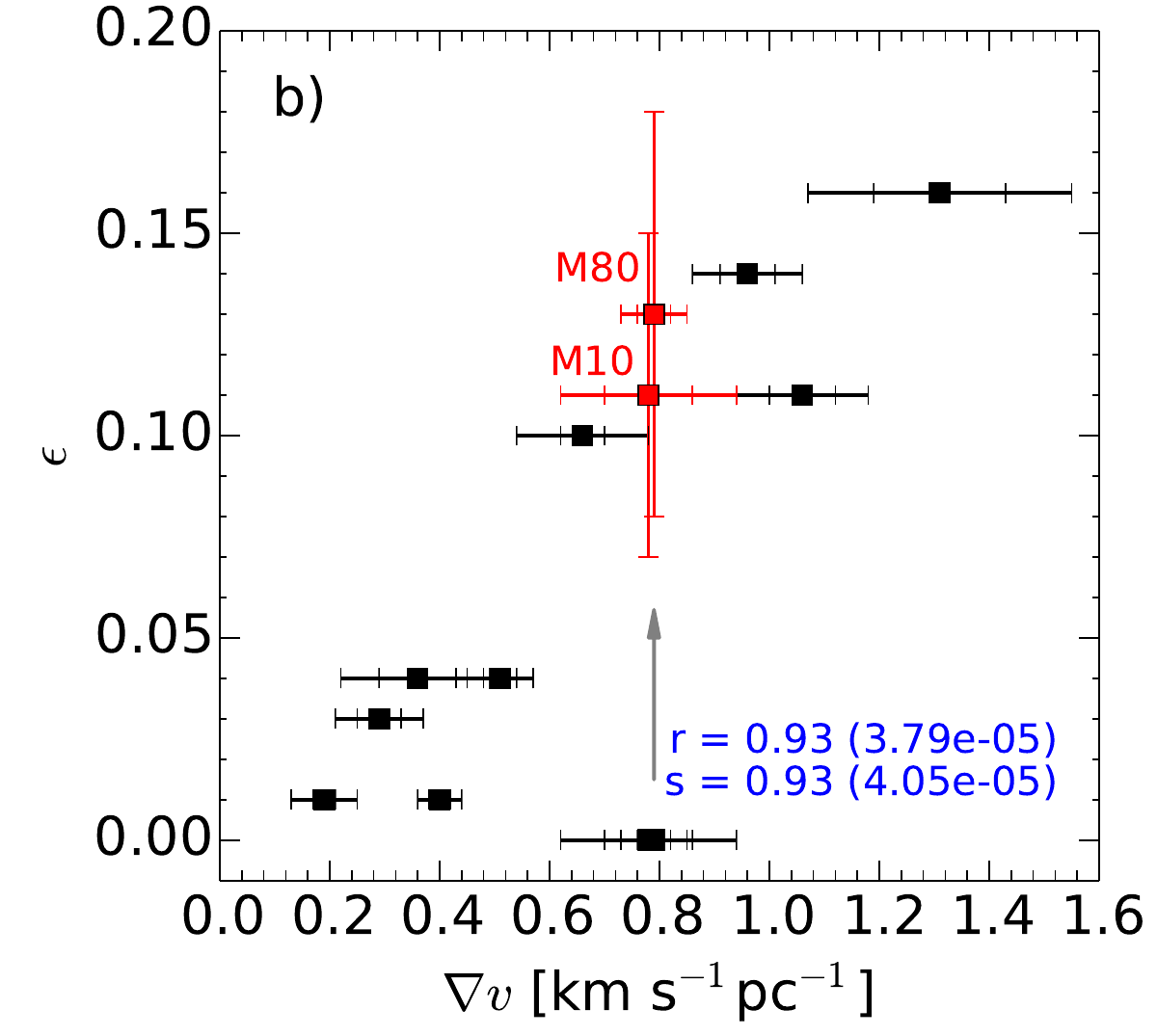}\\
\includegraphics[width=0.4\textwidth]{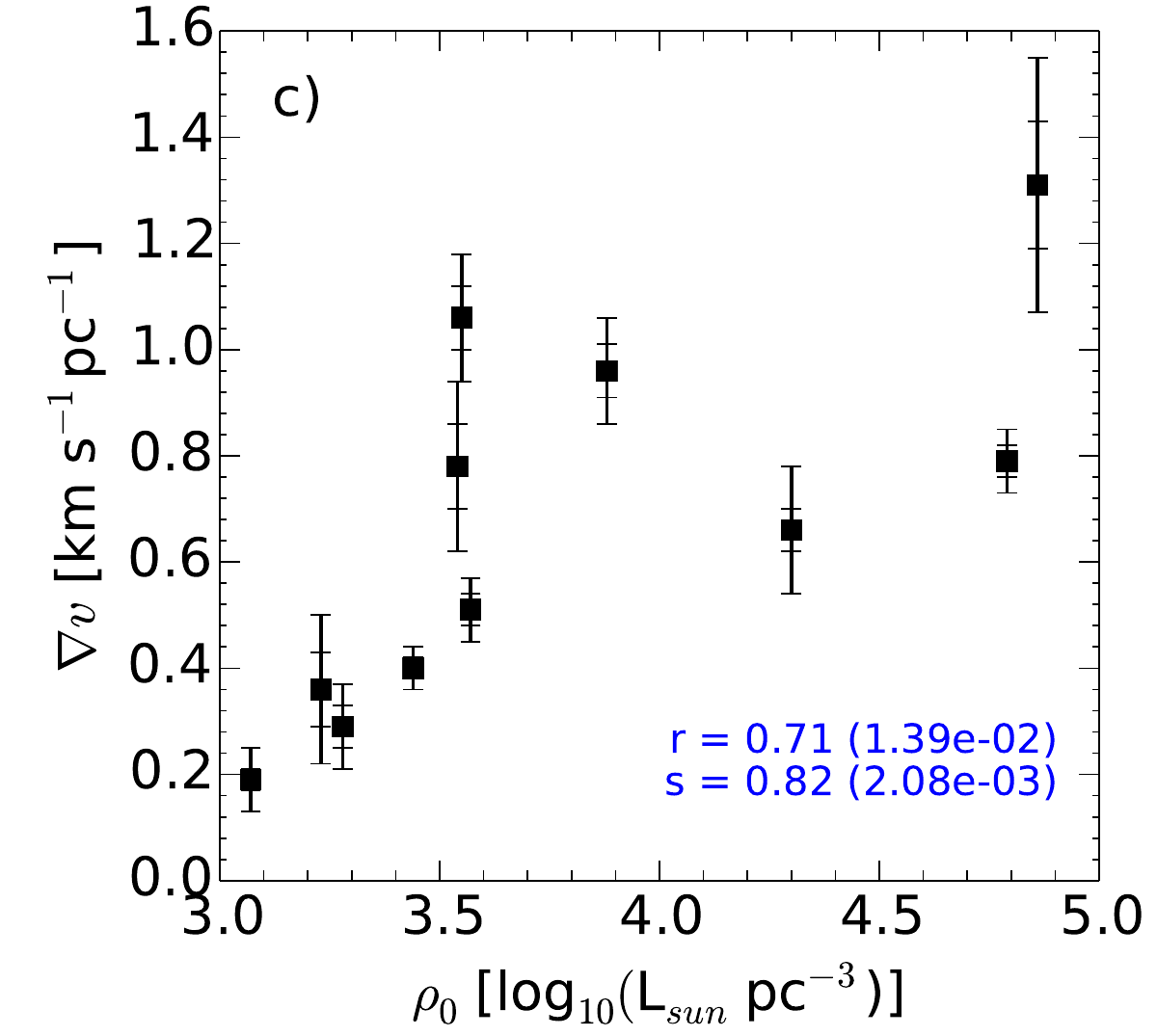}
\includegraphics[width=0.4\textwidth]{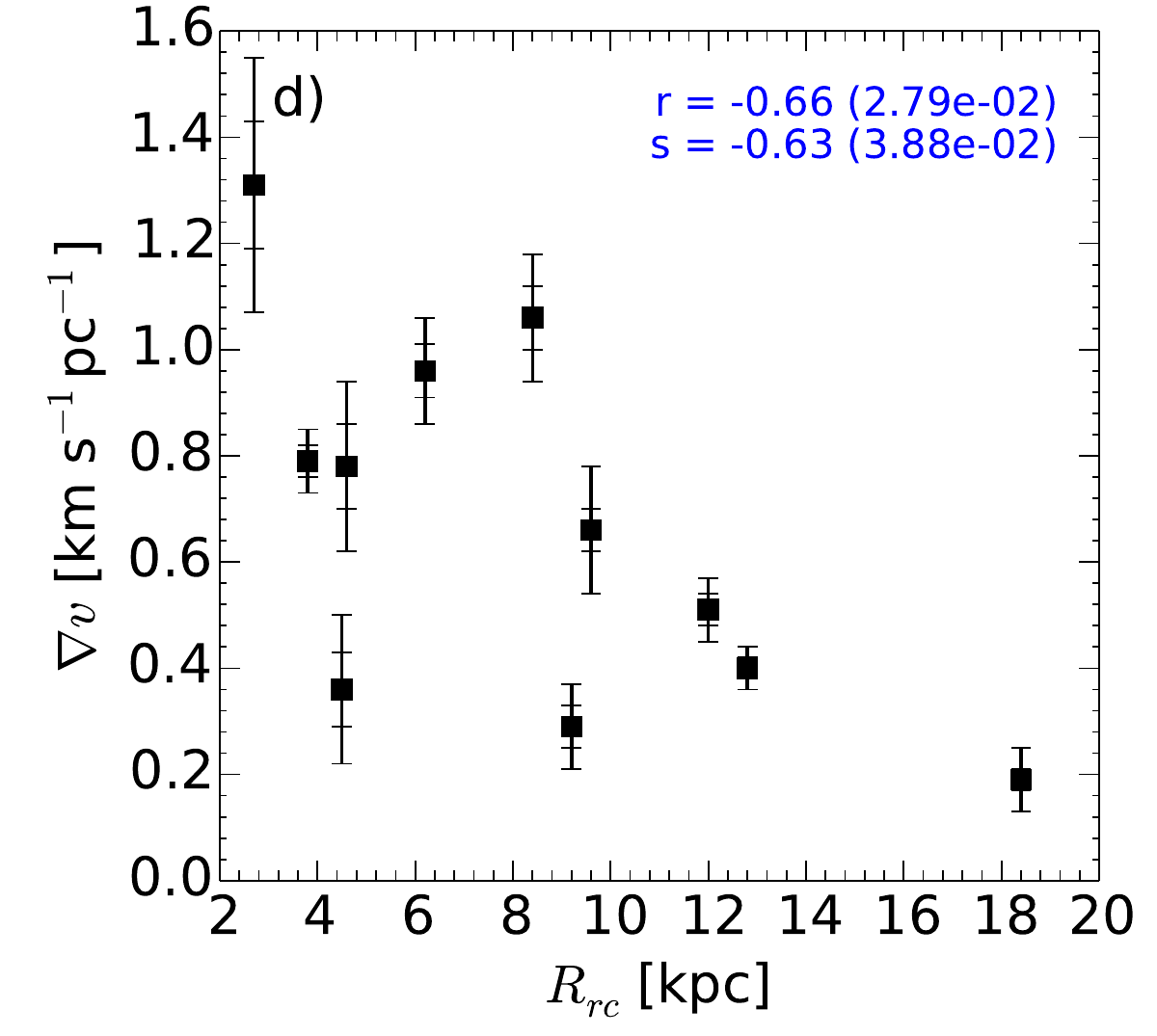}
\end{center}
\caption{
Correlations with parameters from the H96 catalog. {\it a)} central velocity
gradient $\nabla v$ as function of the central velocity dispersion from H96.
{\it b)} ellipticity as function of $\nabla v$. For the two clusters M10 and
M80 we plot the ellipticities from \citet{Chen2010} with red symbols. {\it c)}
$\nabla v$ plotted as a function of the central luminosity density. {\it d)}
$\nabla v$ plotted as function of the distance from the Galactic center.\\ The
Pearson correlation coefficient $r$ and the Spearman's rank correlation $s$ are
shown in all panels. The corresponding two-tailed p-values are given in the
parenthesis. For the ellipticity correlation $r$ and $s$ are computed with the
updated values for M10 and M80 from \citet{Chen2010}.
 }
\label{fig:correlations}
\end{figure*}

As further verification of our derived position angles, we generate synthetic
images from the catalogs by assigning the same magnitude to the stars in the
catalog and adding them to an empty image with a point-spread-function obtained
from the ACS image itself. We then apply a gaussian smoothing of 20\arcsec. We
overplot contours and verify visually that the eigenvector based position
angles are not affected by artefacts in the catalogs.

We estimate the statistical uncertainty using the bootstrapping method in the
same manner as we do for the kinematic datasets. The bin maps show that
artefacts in the catalogs can affect our measurements of the ellipticity. We
therefore run our analysis on hemispheres by mirroring all points of one
hemisphere about the center onto the other side. We repeat this on 36 different
hemispheres, separated by 10 degrees. We report the total spread of values for
the position angle and the ellipticity as systematic error.

The eigenvalue derived ellipticities show no significant values different from
zero inside the VIRUS-W field of view, and consequently, the derived position
angles are poorly constrained. The strongest signal is typically found in the
75\arcsec\ -- 100\arcsec\ radial range.
 
M28 is not part of the HST ACS survey, we therefore construct a catalog using
DAOPHOT \citep{Stetson1987} on an archival HST ACS F625W image (PI Jonathan
Grindlay; see \citealp{Noyola2006} for a description of the process).  We fill
the gaps between the ACS chips and extend the coverage of the catalog in radial
range using WFC3/UVIS F390W (PI Francesco Ferraro) and WFPC2 F555W (PI Roberto
Buonanno) images. The final catalog extends to a radius of 85\arcsec\ from
the cluster center.

\section{Results}
\label{sec:results}
\begin{table*}
\scriptsize
\begin{center}
\caption{Central rotation gradients, kinematic and photometric position angles, and
ellipticities}
\begin{tabular}{lllllllll}
\hline
Identifier	&$t_{obs,total}$& RA			& DEC 			&  $\|\nabla v\|$ 			  				& $PA_{kin}$ 	& $PA_{phot}$	& $100 \times \epsilon_{phot}^{r < 100\arcsec}$\\
			& [h] 				& [J2000] 		& [J2000]		&	[km\,s$^{-1}$ arcmin$^{-1}$] 			& [Deg.]		&  [Deg. ] 		& \\			
	(1)		& 		(2)		&	(3)			& 		(4)		&		(5)		&		(6)					 			&		(7)		&		(8)		\\			
\hline
M3          &1.7		   & 13:42:11.598    & +28:22:37.94  & 1.5 $\pm$ 0.1 $\pm$ 0.1 & 102.2   $\pm$ 4.5   $\pm$ 11.8 & 79.8 $ \pm$ 2.8    $\pm$ 18.0  & 2.2 $\pm$ 0.2 $\pm$ 1.1\\
M5          &3.3		   & 15:18:33.143    & +02:04:52.22  & 2.1 $\pm$ 0.1 $\pm$ 0.1 & 58.5    $\pm$ 2.8   $\pm$  5.6 & 57.6   $\pm$ 4.9   $\pm$ 12.4  & 1.8 $\pm$ 0.3 $\pm$ 0.7\\
M10         &3.5		   & 16:57:08.981    & -04:06:00.44  & 1.0 $\pm$ 0.1 $\pm$ 0.1 & 63.5    $\pm$ 9.0   $\pm$ 14.7 & 111.1  $\pm$ 48.9  $\pm$ 38.7  & 0.6 $\pm$ 0.3 $\pm$ 0.5\\
M12         &5.3		   & 16:47:14.190    & -01:56:53.36  & 0.5 $\pm$ 0.1 $\pm$ 0.1 & 178.9   $\pm$ 9.9   $\pm$ 19.3 & 115.8  $\pm$ 48.8  $\pm$ 83.8  & 0.4 $\pm$ 0.4 $\pm$ 0.7\\
M13         &2.7		   & 16:41:41.147    & +36:27:36.62  & 1.7 $\pm$ 0.1 $\pm$ 0.1 & 106.5   $\pm$ 3.6   $\pm$  7.8 & 115.8  $\pm$ 3.8   $\pm$ 16.8  & 1.8 $\pm$ 0.3 $\pm$ 1.1\\
M28         &2.2		   & 18:24:32.878    & -24:52:13.74  & 2.1 $\pm$ 0.2 $\pm$ 0.2 & 46.4    $\pm$ 6.5   $\pm$ 16.7 & 57.5   $\pm$ 10.3  $\pm$ 34.9  & 1.4 $\pm$ 0.6 $\pm$ 3.0\\
M53         &3.7		   & 13:12:55.208    & +18:10:06.22  & 1.0 $\pm$ 0.2 $\pm$ 0.1 & 22.5    $\pm$ 12.6  $\pm$ 19.2 & 54.4   $\pm$ 4.2   $\pm$ 10.4  & 1.3 $\pm$ 0.2 $\pm$ 0.6\\
M56         &5.3		   & 19:16:35.630    & +30:11:01.44  & 0.8 $\pm$ 0.1 $\pm$ 0.1 & 34.9    $\pm$ 14.7  $\pm$ 19.5 & 69.5   $\pm$ 7.5   $\pm$ 11.6  & 1.4 $\pm$ 0.3 $\pm$ 0.5\\
M80         &2.7		   & 16:17:02.403    & -22:58:34.12  & 2.3 $\pm$ 0.1 $\pm$ 0.1 & 139.0   $\pm$ 3.7   $\pm$  8.5 & 155.4  $\pm$ 17.5  $\pm$ 20.7  & 0.9 $\pm$ 0.2 $\pm$ 0.5\\
M92         &2.7		   & 17:17:07.383    & +43:08:09.23  & 1.6 $\pm$ 0.2 $\pm$ 0.1 & 8.9     $\pm$ 5.0   $\pm$ 12.0 & 55.3   $\pm$ 4.7   $\pm$ 4.4   & 1.7 $\pm$ 0.2 $\pm$ 0.5\\
NGC\,6934   &5.3		   & 20:34:11.346    & +07:24:16.95  & 1.8 $\pm$ 0.1 $\pm$ 0.1 & 170.9   $\pm$ 3.9   $\pm$ 11.9 & 144.9  $\pm$ 66.2  $\pm$ 84.7  & 0.2 $\pm$ 0.3 $\pm$ 0.5\\

\hline
\label{tab:results}
\end{tabular}
\begin{minipage}{.95\textwidth}
Notes-- Column 2 lists the total exposure time (including sky nods) per object.
Columns 3 and 4 list the adopted RA and Dec.\ center positions that we refit as
described in the text. They are given in the reference system of the ACS GC
survey which in turn is referenced against the 2MASS catalog.  Column 5 lists
the absolute value of the central velocity gradient and the statistical errors
(first) and systematic errors (second). Column 6 gives the kinematic position
angle and the corresponding statistical and systematic errors. All angles are
measured from north to east. Columns 7 and 8 give the photometric position
angles and the ellipticities that we derive from the catalogs through the
eigenvector analysis. For both quantities we also list the statistical and
systematic errors. As M28 has no catalog available from the ACS Survey, our
measurements of the position angle and the flattening are based on a catalog
that we compile from HST archival data as described in the text.
\end{minipage}
\end{center}
\end{table*}

Fig.\,\ref{fig:vmaps_ellmaps} shows maps of the line-of-sight velocity for the
11 clusters in our sample. The rotation is clearly seen as a velocity gradient
across the field. We detect statistically significant central rotation in all
clusters. 

We translate the projected absolute values of the velocity gradients to
physical units using distance estimates, core radii, and half light radii from
H96. We find that the velocity gradients have absolute values in the range of
0.2 -- 1.3\,km\,s$^{-1}\,\mathrm{pc}^{-1}$ with average values of
0.7\,km\,s$^{-1}\,\mathrm{pc}^{-1}$, 0.6\,km\,s$^{-1}{r_c}^{-1}$, and 
2.2\,km\,s$^{-1}{r_h}^{-1}$.

In Fig.\,\ref{fig:vmaps_ellmaps} we indicate the derived kinematic position
angles and also overplot the eigenvector derived position angles. The low
ellipticities of M10, M12 and NGC\,6934 leave the values from our eigenvalue
analysis poorly constrained. In the other cases, both position angle
measurements agree well within the errors with one notable exception: for M92
the kinematic position angle deviates from the catalog derived value by about
45\Deg. Unfortunately, in the case of M92 the southern field covered a
relatively sparse region of the cluster and thus, it poorly constrains the
direction of the velocity gradient. We give the absolute values of the central
velocity gradients, the kinematic and the eigenvector based position angles,
adopted cluster centers and the total exposure times in
Table\,\ref{tab:results}.

In Fig.\,\ref{fig:correlations} we compare the absolute values of the central
velocity gradients with other parameters from the H96 catalog. We find a very
tight correlation with the central velocity dispersion, with the exception of
M80. This correlation is possibly the result of dispersion measurements that
include (and do not correct for) rotation. For example, M13 has the second
largest published value of the central velocity dispersion in our sample
\citep{Cohen2005}. By extrapolating the central rotation gradient to the actual
positions of the stars we can derive a velocity dispersion of
9.2\,km\,s$^{-1}$. This extrapolation does not take the flattening of the
rotation curve into account and therefore it is not surprising that this value is
actually larger than the \citet{Cohen2005} value of 7.1\,km\,s$^{-1}$. It does
show, however, that rotation can significantly impact the dispersion
measurements. For M80, the rotation can probably not explain the large value of
the central dispersion of 12.4\,km\,s$^{-1}$ \citep{Dubath1997}.

The observed correlation between the outer ellipticity and the central rotation
however is most likely physical.  The only outliers (M10 and M80) fall nicely
on the correlation when the ellipticity values from H96 are replaced by those
of \citet{Chen2010}.
  
We also find a correlation between central luminosity density and the value of
the velocity gradient. More speculative is a possible correlation of the amount
of central rotation with the distance from the galactic center as shown in the
lower right panel. If the flattening does increase towards the galactic center
as suggested by \citet{White1987} and \citet{Chen2010} and the flattening is as
closely related to rotation as we find, then this may be expected.
 
\label{sec:conclusions}
\begin{figure}[h!]
\begin{center}
\includegraphics[width=0.4\textwidth]{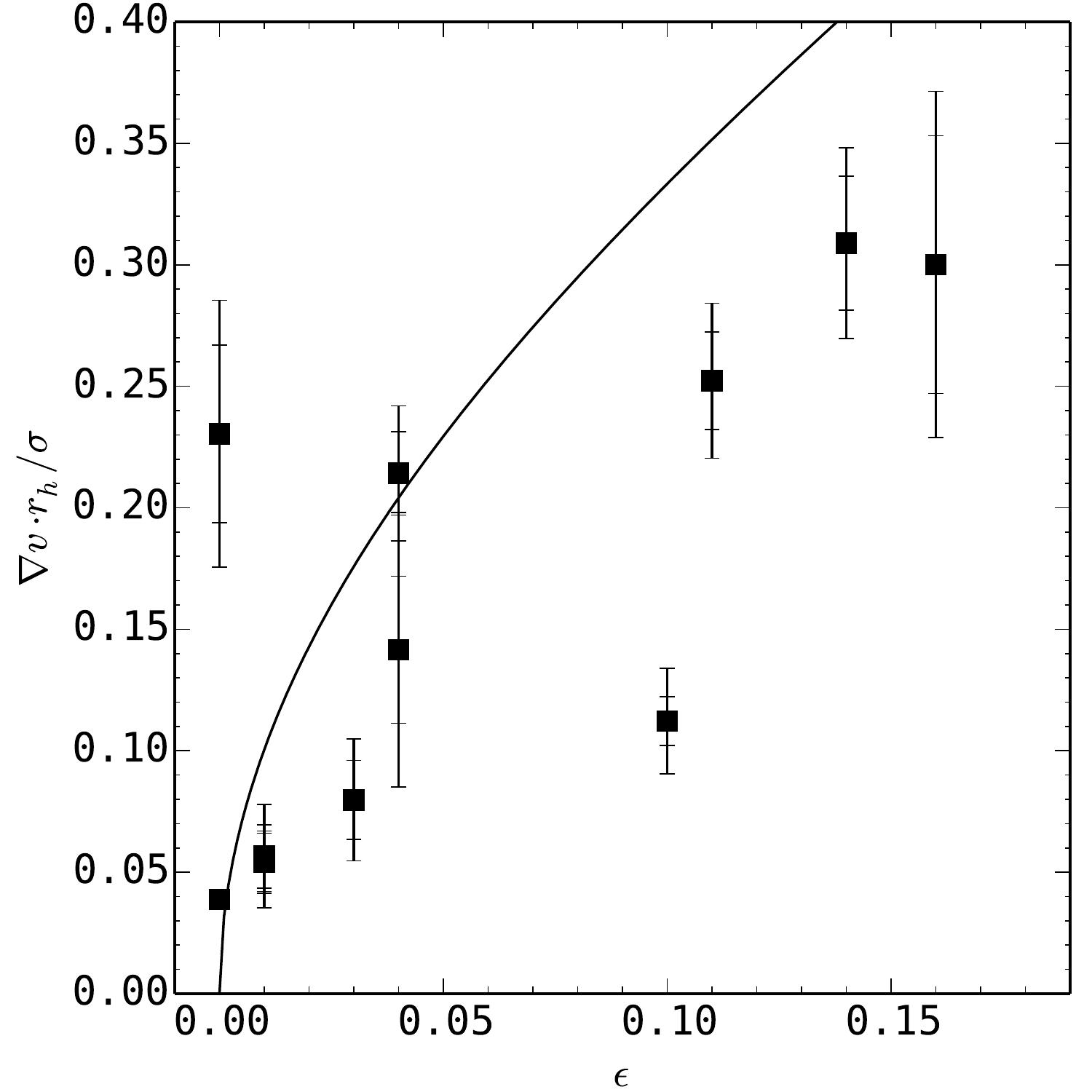}
\end{center}
\caption{
(V/$\sigma$,$\epsilon$) diagram for the GCs in our sample. As proxy for the
maximum rotational velocity we multiply our derived velocity gradients by the
half-light radii from H96. The ellipticities are also taken from the same
catalog. The solid line shows the location for isotropic rotators
\citep{Binney1987}. Shorter error bars indicate the systematic errors, longer
error bars the systematic plus the random errors.
}
\label{fig:v_sigma}
\end{figure}

With estimates for ellipticities, rotation, and velocity dispersion, we plot
our objects in the (V/$\sigma$,$\epsilon$) diagram (Fig.\,\ref{fig:v_sigma}).
The values for the rotation are obtained by multiplying the value of the
central velocity gradient by the half-light radius $r_h$ from H96. The solid
line indicates the locus of edge-on isotropic oblate rotators
\citep{Binney1987}. All but two of our objects fall below this line. However,
as discussed before, the estimates for the central dispersion are likely
affected by rotation and also the extrapolation of the rotational velocity to
$r_h$ may yield incorrect values for the actual maximum rotation. On the other
hand our analysis does not take projection effects into accout which may move
objects even further away from the line. While we have no reason to expect that
GCs behave like isotropic oblate rotators, the clusters in our sample follow a
general trend of larger (V/$\sigma$) for increased ellipticity which again
indicates that rotation is the driving factor for the flattening.

\section{Conclusions}

We present central velocity fields for 11 Milky Way GCs derived from data
collected in an ongoing survey using the fiber based, optical IFU instrument
VIRUS-W. We find that all clusters presented here show significant gradients in
the observed velocity fields that are indicative of rotation. 

This is surprising and clearly shows a need for theoretical models to produce
and to sustain rotation on scales comparable to the core radius. We also find that
the orientation of the central flattening in stellar density (within an
100\arcsec\ aperture) is generally in good agreement with the kinematic
position angle that we derive from the velocity fields.
 
We show that the central rotation correlates very well with published values
for the central velocity dispersion, thus pointing to a possible impact from
the rotation. A detailed analysis of this effect and a comparison with
dispersion estimates from our data will be included in a subsequent
publication. We also find a strong correlation with outer ellipticity
indicating that, at least for the objects in our sample, the flattening is
primarily due to rotation rather than the Milky Way tidal field.

There is an indication of an increase of rotation towards the galactic center.
If true, this might have a strong impact on formation theories and clearly
shows the need for a larger sample.

One caveat of our study so far is that our sample so far lacks any core
collapsed clusters. It is conceivable that this process will eradicate rotation
through the transport of angular momentum to the outskirts of the cluster.
Future observations will specifically target such systems. Our survey of 27\
clusters will allow us to probe in detail how rotations correlate with
properties such as central velocity dispersions, ellipticity, ages, spatial
distribution, and total luminosity. This will complement both upcoming higher
spatial resolution studies with instruments such as PMAS at the Calar Alto
3.5\,m \citep{Kamann2014} and MUSE at the VLT, and multi-object spectroscopic
studies that are limited to much larger radii. Our derived position angles will
allow the multi-object studies to specifically target angular ranges of maximum
rotation.

\section*{Acknowledgements}
We thank the anonymous referee for the thoughtful comments.  This research has
made use of the NASA/IPAC Extragalactic Database (NED) which is operated by the
Jet Propulsion Laboratory, California Institute of Technology, under contract
with the National Aeronautics and Space Administration. Some of the data
presented in this paper were obtained from the Mikulski Archive for Space
Telescopes (MAST). STScI is operated by the Association of Universities for
Research in Astronomy, Inc., under NASA contract NAS5-26555. Support for MAST
for non-HST data is provided by the NASA Office of Space Science via grant
NNX13AC07G and by other grants.
\bibliographystyle{apj}

\begin{thebibliography}{0}
\expandafter\ifx\csname natexlab\endcsname\relax\def\natexlab#1{#1}\fi

\end{thebibliography}


\begin{thebibliography}{31}

\bibitem[{{Anderson} \& {King}(2003)}]{Anderson2003}
{Anderson}, J. \& {King}, I.~R. 2003, \aj, 126, 772

\bibitem[{{Bekki}(2010)}]{Bekki2010}
{Bekki}, K. 2010, \apjl, 724, L99

\bibitem[{{Bellazzini} {et~al.}(2012){Bellazzini}, {Bragaglia}, {Carretta},
  {Gratton}, {Lucatello}, {Catanzaro}, \& {Leone}}]{Bellazzini2012}
{Bellazzini}, M., {Bragaglia}, A., {Carretta}, E., {Gratton}, R.~G.,
  {Lucatello}, S., {Catanzaro}, G., \& {Leone}, F. 2012, \aap, 538, A18

\bibitem[{{Bianchini} {et~al.}(2013){Bianchini}, {Varri}, {Bertin}, \&
  {Zocchi}}]{Bianchini2013}
{Bianchini}, P., {Varri}, A.~L., {Bertin}, G., \& {Zocchi}, A. 2013, \apj, 772,
  67

\bibitem[{{Binney} \& {Tremaine}(1987)}]{Binney1987}
{Binney}, J. \& {Tremaine}, S. 1987, {Galactic dynamics}

\bibitem[{{Chen} \& {Chen}(2010)}]{Chen2010}
{Chen}, C.~W. \& {Chen}, W.~P. 2010, \apj, 721, 1790

\bibitem[{{Cohen} \& {Mel{\'e}ndez}(2005)}]{Cohen2005}
{Cohen}, J.~G. \& {Mel{\'e}ndez}, J. 2005, \aj, 129, 303

\bibitem[{{Dubath} {et~al.}(1997){Dubath}, {Meylan}, \& {Mayor}}]{Dubath1997}
{Dubath}, P., {Meylan}, G., \& {Mayor}, M. 1997, \aap, 324, 505

\bibitem[{{Einsel} \& {Spurzem}(1999)}]{Einsel1999}
{Einsel}, C. \& {Spurzem}, R. 1999, \mnras, 302, 81

\bibitem[{{Ernst} {et~al.}(2007){Ernst}, {Glaschke}, {Fiestas}, {Just}, \&
  {Spurzem}}]{Ernst2007}
{Ernst}, A., {Glaschke}, P., {Fiestas}, J., {Just}, A., \& {Spurzem}, R. 2007,
  \mnras, 377, 465

\bibitem[{{Fabricius} {et~al.}(2014){Fabricius}, {Coccato}, {Bender}, {Drory},
  {Goessl}, {Landriau}, {Saglia}, {Thomas}, \& {Williams}}]{Fabricius2014}
{Fabricius}, M.~H., {Coccato}, L., {Bender}, R., {Drory}, N., {Goessl}, C.,
  {Landriau}, M., {Saglia}, R.~P., {Thomas}, J., \& {Williams}, M.~J. 2014,
  ArXiv e-prints

\bibitem[{{Fabricius} {et~al.}(2012){Fabricius}, {Grupp}, {Bender}, {Drory},
  {Arns}, {Barnes}, {G{\"o}ssl}, {Snigula}, {Hill}, {Hopp}, {Lang-Bardl},
  {MacQueen}, {Saglia}, \& {Wullstein}}]{Fabricius2012b}
{Fabricius}, M.~H., {Grupp}, F., {Bender}, R., {Drory}, N., {Arns}, J.,
  {Barnes}, S., {G{\"o}ssl}, C., {Snigula}, J., {Hill}, G.~J., {Hopp}, U.,
  {Lang-Bardl}, F., {MacQueen}, P.~J., {Saglia}, R., \& {Wullstein}, P. 2012,
  in Society of Photo-Optical Instrumentation Engineers (SPIE) Conference
  Series, Vol. 8446, Society of Photo-Optical Instrumentation Engineers (SPIE)
  Conference Series

\bibitem[{{Gebhardt} {et~al.}(2000){Gebhardt}, {Richstone}, {Kormendy},
  {Lauer}, {Ajhar}, {Bender}, {Dressler}, {Faber}, {Grillmair}, {Magorrian}, \&
  {Tremaine}}]{Gebhardt2000b}
{Gebhardt}, K., {Richstone}, D., {Kormendy}, J., {Lauer}, T.~R., {Ajhar},
  E.~A., {Bender}, R., {Dressler}, A., {Faber}, S.~M., {Grillmair}, C.,
  {Magorrian}, J., \& {Tremaine}, S. 2000, \aj, 119, 1157

\bibitem[{{Geyer} {et~al.}(1983){Geyer}, {Nelles}, \& {Hopp}}]{Geyer1983}
{Geyer}, E.~H., {Nelles}, B., \& {Hopp}, U. 1983, \aap, 125, 359

\bibitem[{{G{\"o}ssl} \& {Riffeser}(2002)}]{Gossl2002}
{G{\"o}ssl}, C.~A. \& {Riffeser}, A. 2002, \aap, 381, 1095

\bibitem[{{Harris}(1996)}]{Harris1996}
{Harris}, W.~E. 1996, \aj, 112, 1487

\bibitem[{{Hill} {et~al.}(2004){Hill}, {Gebhardt}, {Komatsu}, \&
  {MacQueen}}]{Hill2004}
{Hill}, G.~J., {Gebhardt}, K., {Komatsu}, E., \& {MacQueen}, P.~J. 2004, in
  American Institute of Physics Conference Series, Vol. 743, The New Cosmology:
  Conference on Strings and Cosmology, ed. {R.~E.~Allen, D.~V.~Nanopoulos, \&
  C.~N.~Pope}, 224--233

\bibitem[{{Kamann} {et~al.}(2014){Kamann}, {Wisotzki}, {Roth}, {Gerssen},
  {Husser}, {Sandin}, \& {Weilbacher}}]{Kamann2014}
{Kamann}, S., {Wisotzki}, L., {Roth}, M.~M., {Gerssen}, J., {Husser}, T.-O.,
  {Sandin}, C., \& {Weilbacher}, P. 2014, ArXiv e-prints

\bibitem[{{Lane} {et~al.}(2011){Lane}, {Kiss}, {Lewis}, {Ibata}, {Siebert},
  {Bedding}, {Sz{\'e}kely}, \& {Szab{\'o}}}]{Lane2011}
{Lane}, R.~R., {Kiss}, L.~L., {Lewis}, G.~F., {Ibata}, R.~A., {Siebert}, A.,
  {Bedding}, T.~R., {Sz{\'e}kely}, P., \& {Szab{\'o}}, G.~M. 2011, \aap, 530,
  A31

\bibitem[{{Longaretti} \& {Lagoute}(1997)}]{Longaretti1997}
{Longaretti}, P.-Y. \& {Lagoute}, C. 1997, \aap, 319, 839

\bibitem[{{Mastrobuono-Battisti} \& {Perets}(2013)}]{Mastrobuono-Battisti2013}
{Mastrobuono-Battisti}, A. \& {Perets}, H.~B. 2013, \apj, 779, 85

\bibitem[{{Meylan} \& {Heggie}(1997)}]{Meylan1997}
{Meylan}, G. \& {Heggie}, D.~C. 1997, \aapr, 8, 1

\bibitem[{{Mor{\'e}} {et~al.}(1980){Mor{\'e}}, {Garbow}, \&
  {Hillstrom}}]{More1980}
{Mor{\'e}}, J.~J., {Garbow}, B.~S., \& {Hillstrom}, K.~E. 1980, User Guide for
  MINPACK-1

\bibitem[{{Noyola} \& {Gebhardt}(2006)}]{Noyola2006}
{Noyola}, E. \& {Gebhardt}, K. 2006, \aj, 132, 447

\bibitem[{{Sarajedini} {et~al.}(2007){Sarajedini}, {Bedin}, {Chaboyer},
  {Dotter}, {Siegel}, {Anderson}, {Aparicio}, {King}, {Majewski},
  {Mar{\'{\i}}n-Franch}, {Piotto}, {Reid}, \& {Rosenberg}}]{Sarajedini2007}
{Sarajedini}, A., {Bedin}, L.~R., {Chaboyer}, B., {Dotter}, A., {Siegel}, M.,
  {Anderson}, J., {Aparicio}, A., {King}, I., {Majewski}, S.,
  {Mar{\'{\i}}n-Franch}, A., {Piotto}, G., {Reid}, I.~N., \& {Rosenberg}, A.
  2007, \aj, 133, 1658

\bibitem[{{Stetson}(1987)}]{Stetson1987}
{Stetson}, P.~B. 1987, \pasp, 99, 191

\bibitem[{{Trager} {et~al.}(1995){Trager}, {King}, \&
  {Djorgovski}}]{Trager1995}
{Trager}, S.~C., {King}, I.~R., \& {Djorgovski}, S. 1995, \aj, 109, 218

\bibitem[{{van den Bosch} {et~al.}(2006){van den Bosch}, {de Zeeuw},
  {Gebhardt}, {Noyola}, \& {van de Ven}}]{van-den-Bosch2006}
{van den Bosch}, R., {de Zeeuw}, T., {Gebhardt}, K., {Noyola}, E., \& {van de
  Ven}, G. 2006, \apj, 641, 852

\bibitem[{{van Leeuwen} \& {Le Poole}(2002)}]{van-Leeuwen2002}
{van Leeuwen}, F. \& {Le Poole}, R.~S. 2002, in Astronomical Society of the
  Pacific Conference Series, Vol. 265, Omega Centauri, A Unique Window into
  Astrophysics, ed. F.~{van Leeuwen}, J.~D. {Hughes}, \& G.~{Piotto}, 41

\bibitem[{{White} \& {Shawl}(1987)}]{White1987}
{White}, R.~E. \& {Shawl}, S.~J. 1987, \apj, 317, 246

\bibitem[{{Zocchi} {et~al.}(2012){Zocchi}, {Bertin}, \& {Varri}}]{Zocchi2012}
{Zocchi}, A., {Bertin}, G., \& {Varri}, A.~L. 2012, \aap, 539, A65

\end{thebibliography}

\end{document}